# Surface transport properties of Fe-based superconductors: The influence of degradation and inhomogeneity

T. Plecenik,[1] M. Gregor,[1] R. Sobota,[1] M. Truchly,[1] L. Satrapinskyy,[1] F. Kurth,[2] B. Holzapfel,[2] K. Iida,[2] P. Kus,[1] and A. Plecenik[1]

[1]*Department of Experimental Physics, FMPI, Comenius University, 842 48 Bratislava, Slovakia*
[2]*Institute for Metallic Materials, IFW Dresden, P. O. Box 270116, D-01171 Dresden, Germany*

Surface properties of Co-doped $BaFe_2As_2$ epitaxial superconducting thin films were inspected by X-ray photoelectron spectroscopy, scanning spreading resistance microscopy (SSRM), and point contact spectroscopy (PCS). It has been shown that surface of Fe-based superconductors degrades rapidly if being exposed to air, what results in suppression of gap-like structure on PCS spectra. Moreover, SSRM measurements revealed inhomogeneous surface conductivity, what is consistent with strong dependence of PCS spectra on contact position. Presented results suggest that fresh surface and small probing area should be assured for surface sensitive measurements like PCS to obtain intrinsic properties of Fe-based superconductors.

Since the discovery of Fe-based superconductors, great effort has been made to determine their fundamental properties such as width of energy gaps and pairing symmetry. Experimental results obtained by different groups and methods on various sample forms are however inconsistent. For instance, shapes of the differential conductance spectra obtained by point contact spectroscopy (PCS) and width of energy gaps extracted from them for Co-doped $BaFe_2As_2$ (Ba-122) are varying from one measurement to another, and the results are interpreted within single gap or multi-gap models.[1-5] The pairing symmetry in Fe-based superconductors in general is also still under dispute.[6] A possible fundamental reason for such discrepancy is that the surface of Co-doped Ba-122 is highly sensitive against ambient air. Point contacts as well as other types of junctions on Fe-based superconductors are often prepared and measured *ex-situ*, which leads to undefined interface of such junctions and varying experimental results. It should be noted that recently suggested phase-sensitive experiments designed to test a pairing symmetry in Fe-based superconductors[7] are also based on point contacts and tunnel junctions, which are both highly surface-sensitive. Undefined surfaces and interfaces are also a major drawback for fabrication of various cryo-electronic devices such as Josephson junctions and SQUIDs, although short coherence length remains the main problem in this material class.

Here we demonstrate how the surface of Co-doped Ba-122 film is degraded by measuring X-ray photoelectron spectroscopy (XPS) on three different types of films; (i) as-received (i.e., stored in a plastic bag filled with Ar gas and transferred to the XPS measurements), (ii) Ar ion beam etching (IBE) processed films, and (iii) exposed to ambient air after IBE process. Additionally, by scanning spreading resistance microscopy (SSRM) measurements, we found inhomogeneity in surface conductivity for all samples that is also responsible for inconsistent results of surface sensitive experiments. PCS measurements on the aforementioned films confirmed a gap-like structure only for as-received and IBE processed films, whereas degraded films (i.e., exposed to air for more than one hour after IBE) do not show such structure.

Additionally, PCS spectra strongly depend on the contact position, which is consistent with the results of SSRM measurements. Measurements of time evolution of point contact resistance also show that even contacts prepared on clean surfaces degrade if being exposed to air, most likely due to oxygen diffusion along the interface.

The epitaxial Co-doped Ba-122 superconducting thin films on $CaF_2$ (001) substrates were prepared by pulsed laser deposition (PLD), where the $Ba(Fe_{0.9}Co_{0.1})As_2$ target was ablated with 248 nm KrF radiation at a frequency of 7 Hz under UHV conditions (base pressure $10^{-9}$ mbar). A deposition temperature of 700 °C was employed. The detailed preparation procedure can be found in Ref. 8. After the deposition samples were taken out from the UHV chamber and kept in a plastic bag filled in protective Ar atmosphere. It is note that films were exposed to air during this procedure for a short period of time. All samples were unpacked from the plastic bags immediately before measurements. The X-ray photoelectron spectra were recorded on Omicron multiprobe system with hemispherical analyzer and monochromatic Al Kα X-rays (1486.6 eV). Spectra were measured at ambient temperature with photoemission of 45 °C from the surface. To minimize the effects of charging on the Co-doped Ba-122, a low-energy (below 1 eV) electron gun was used for charge neutralization. The carbon 1s line (for hydrocarbon, binding energy 284.8 eV) has been used to calibrate the binding-energy scale for XPS measurements. The IBE has been done in the same vacuum chamber. The atomic force microscopy (AFM) and SSRM measurements were done by NT-MDT NTegra Aura microscope in a chamber with pure nitrogen atmosphere. Standard silicon AFM tips with conductive Pt coating were used. During the measurement, the sample was grounded and the tip was biased by voltage, while the current through the tip/sample contact was monitored. For the PCS, the V-shaped PtIr wire with 200 μm diameter providing small point contact area and small force applied to the surface was used. Differential conductance (dI/dV-V) characteristics were measured at a temperature of 4.2 K by four probe, low-frequency phase-sensitive technique using a Stanford Research Systems SR830 lock-in nanovoltmeter,

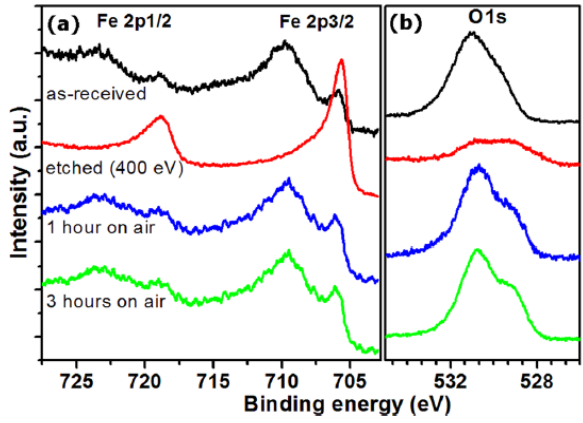

FIG. 1. X-ray photoelectron spectra of (a) Fe 2p and (b) O 1s peaks collected on as prepared sample, after IBE (400eV, 90 min) and after exposition to air for 1 and 3 h (from top to bottom).

Kithley 202 current source, and Keithley 2000 multimeter.

At first, surfaces of the as-received samples were inspected by all methods mentioned above. All samples were then etched by Ar IBE (400 eV ion energy) in the XPS vacuum chamber and the samples for SSRM and PCS were transferred back to respective setups (contact with air was limited to <5 min). All the aforementioned methods were applied again immediately after the etching. In the next step, the samples were exposed to air for 1 h and measured again. The same procedure was applied after additional 2 h of air exposure.

The XPS investigation identified that the surface of as-received film was contaminated by carbon and oxygen. It is note that all photoelectrons peaks related to the Co-doped Ba-122 were visible (Table I). The Co peak is difficult to isolate from the survey spectra due to the fact that Co 2p has almost the same binding energy as Ba 3d. The Fe 2p spectrum in Fig. 1 indicates that Fe is present in more than one chemical state with binding energies of 705.9 and 709.7 eV for Fe 2p3/2. The peak at 705.9 eV is coming from pnictide and is very close to binding energy for Fe metal, while the peak with energy of 709.7eV corresponds to the chemical state $Fe^{2+}$ or $Fe^{3+}$, indicating the presence of FeO or $Fe_2O_3$ mainly on the surface of as-received and air exposed samples.[9] After the IBE, the peak with an energy of 709.7eV disappeared, and only the peak at 705.9 eV was observed. The wide and asymmetric peak of O 1s spectrum indicates that oxygen is present in two chemical states (Fig. 1). It consists of two peaks with binding energies of 530.7 eV, which indicates $Fe_2O_3$ and hydroxyl groups (surface contamination by carbon), and 529 eV, which is attributed to FeO and BaO.[10] After the IBE, the oxygen concentration was considerably reduced, however, returned to the same level after being exposed to air (Fig. 1, Table I). Slightly higher concentration of Ba and residual oxygen visible after the IBE is a consequence of BaO precipitates, which were observed on the surface.

Even though the surface roughness of measured samples was very small (<2 nm peak-to peak, see upper part of Fig. 2), resulting SSRM images show inhomogeneous surface conductivity on nanometer scale in all cases (lower part of Fig. 2). The average current varied with positions on the sample and was unstable in general. The maps of surface conductivity were similar in all cases (i.e., as-received, IBE processed, and air exposed after the IBE) but the maximal current value (z-bar in lower part of Fig. 2) varied from several tens of pA for as-received samples up to several tens of nA on IBE processed surfaces. After being exposed to air, the average current is observed to decrease again to several nA after 1 h of exposure and several hundred pA after additional 2 h (i.e., total 3 h). A variation of the average current among the three different types of films can be explained by the surface degradation processes described above. Most importantly, the high inhomogeneity of the surface conductivity in all cases shows that the transport properties of Co-doped Ba-122 surfaces are highly position-dependent. If a large area is investigated, e.g., a large PCS contact, integral signal is observed. Taking into account also the XPS results, within one junction may coexist several SIN channels with tunneling current $I_T$ SN channels exhibiting the exhibiting the Andreev reflection with current $I_A$, SS'N channels with current $I_D$, SN'N channels etc., where S is non-degraded Co-doped Ba-122 superconductor, S' is Co-doped Ba-122 with suppressed superconductivity, N' is degraded Co-doped Ba-122 interface layer in normal state, I is degraded Co-doped Ba-122 interface layer in insulating state, and N is normal metal (electrode). Accordingly, the total current is expressed in the form $I = \alpha I_T + \beta I_A + \gamma I_D + \ldots$ and the shape of the final I-V and dI/dV-V curves depends on the weight coefficients α, β, γ, etc. One can thus restore differential characteristics exhibiting various or no gap-like structures on the same surface.

TABLE I. Atomic concentration (%) of elements in Co-doped Ba-122 films calculated from the survey XPS spectra of as-received, IBE and air exposed samples. The concentrations do not correspond to the Ba-122 chemical formula as the amount of Ba and O is increased due to existence of BaO precipitates on the surface also after IBE.

|  | O 1s (531 eV) | C 1s (284.5 eV) | As 3p (139 eV) | Ba 4d (89 eV) | Fe 2p (710 eV) |
| --- | --- | --- | --- | --- | --- |
| As-received | 35.3 | 48 | 4 | 9.2 | 3.5 |
| Etched | 19 | 0 | 25.2 | 30.8 | 25 |
| Air/1 h | 38.7 | 35.4 | 7.1 | 12.6 | 6.2 |
| Air/2 h | 36.8 | 41.9 | 6.4 | 10.9 | 4.1 |

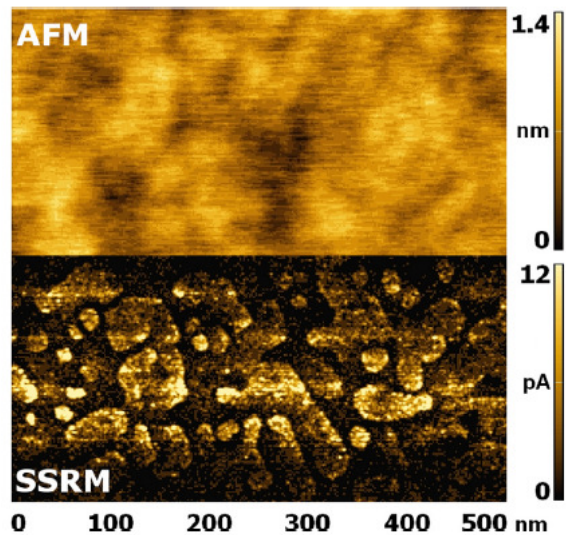

FIG. 2. Typical AFM topography and SSRM map of the surface conductivity measured on as-received samples.

The differential conductance spectra measured by the PCS reflected mostly surface changes indicated by the SSRM and XPS. PtIr/Co-doped Ba-122 point contacts were prepared on different places of the sample. Spectra presented below were representative results. At first we tested as-received surfaces. In accordance with SSRM measurements, two types of differential characteristics— either with or without gap-like structure were measured due to inhomogeneous surface conductivity, both with strong decreasing of differential conductivity with bias voltage. In the first type of the characteristics, two gap-like structures below ±10 mV originating from Andreev reflection are clearly visible (see Fig. 3(a), curve 1 for details). However, no gap-like structures are visible in the second type of the characteristics (see Fig. 3(a), curve 2) measured on the same surface. These results fully reflect SSRM and XPS measurements where inhomogeneous and partially degraded surface layer was detected. As the surface area of point contacts is small, all obtained differential characteristics reflect rather local properties of the Co-doped Ba-122 surface in contrast to those measured on large contacts (e.g., in Ref. 4). After the degraded layer was removed by the Ar IBE, most of the differential characteristics exhibited clear gap-like structure (Fig. 3(b), curve 1), although their shape still varied with the contact position (curves 1 and 2 in Fig. 3(b)). No significant changes of the differential conductance at higher voltages were observed. In contradiction with the spectra measured on the as-received samples, those measured on the etched surfaces exhibited only one gap-like structure in most cases. It is thus impossible to conclude if the two gaps measured in c-axis originate from intrinsic properties of the Co-doped Ba-122, or from creation of parallel conducting channels containing different (degraded and non-degraded) superconducting phases. It should be noted that the damage by ion bombardment can also play a role on the IBE processed surfaces. After exposure to air for 1 h differential conductance once again became a similar shape of the curve 2 from Fig. 3(a) with decreasing differential conductivity with bias voltage. We thus believe that this decreasing of differential conductivity, which is often observed on Fe-based superconductors,[1,4,5] is related to the degraded surface layer. In some cases a gap-like structure at 1 ~ 2 mV was visible (Fig. 3(c), curve 1). This structure could be associated with degraded superconducting phase on the surface. After additional exposure to air for 2 h, the spectra exhibited no gap-like structure but a sharp peak at zero bias appeared.

To further manifest the influence of degradation on PCS spectra, a PtIr/Co-doped Ba-122 point contact was prepared on IBE surface as in previous cases and evolution of its resistance as a function of time was investigated, while whole arrangement was exposed to air. Within 160 h the contact resistance changed from 47.7 to 138.5 Ω, what we believe is caused by the degradation of the junction interface (Fig. 4). Measured curve followed exponential decay behavior with a time constant $τ$ of 59.5 h, which is qualitatively similar to the previously observed degradation of contacts made on high-$T_C$ superconductors.[11,12] Such long degradation time is most likely caused by the fact that the junction interface is not exposed to the atmosphere directly and the oxygen causing the degradation first has to diffuse along the interface and/or junction imperfections towards the center of the junction. Obtained time constant thus probably reflects mostly this oxygen diffusion process. Nevertheless, it is clear that the junction changes strongly in time if exposed to air, what should be taken into account when such junctions are being prepared.

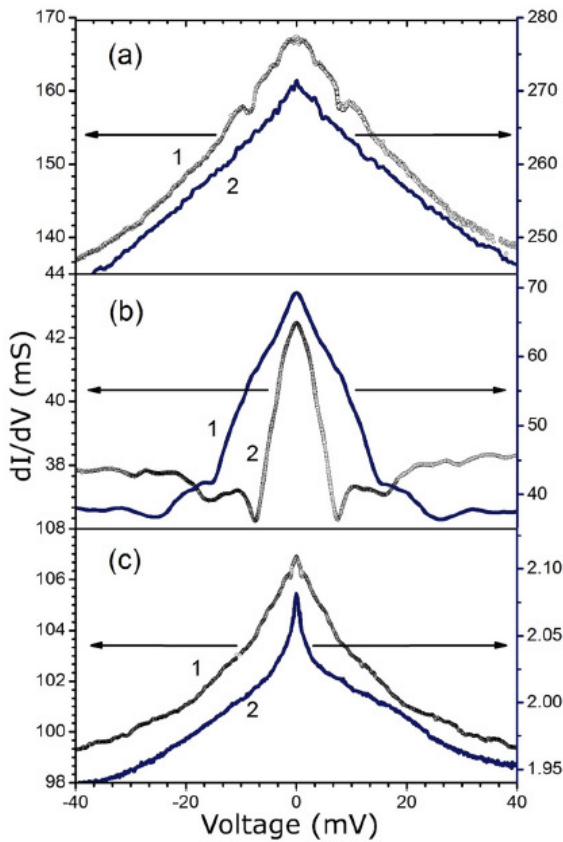

FIG. 3. Typical differential characteristics measured by PCS on (a) as-received Co-doped Ba-122 samples, (b) after ion beam etching, and (c) after etching and exposition to air.

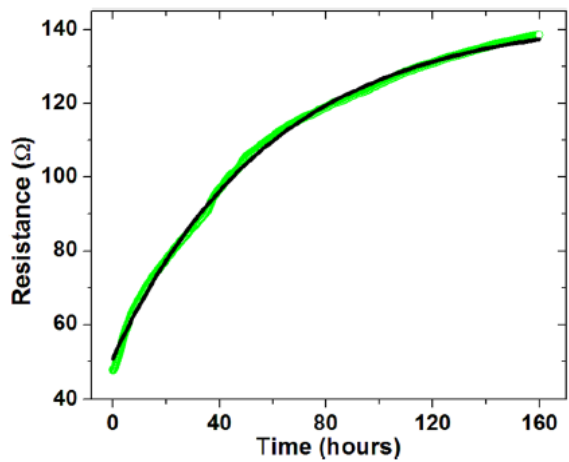

FIG. 4. Time evolution of the PCS junction resistance (green circles) and its exponential decay fit (black solid line). During the measurement the junction was kept in ambient atmosphere at room temperature.

Although all our results were obtained on Co-doped Ba-122 films, we believe that our conclusions are valid also for other Fe-based superconductors as degradation caused by air exposure was reported also in other systems, e.g., FeSe and F-doped LaFeAsO.[13-15] Considering the scanning depth of methods used above (XPS and PCS), which is several nanometers, we can assume that the degradation affects at least several atomic layers. Surface termination peculiarities of various Fe-based systems thus most likely do not play significant role in the degradation process.

To conclude, measurements of superconducting properties of Fe-based superconductors by surface-sensitive methods like PCS are highly influenced by degradation and inhomogeneity of surfaces and interfaces. Obtained results (e.g., gap-like structure in the case of PCS) thus do not have to reflect only the intrinsic properties. This could explain a large variation of reported values of energy gaps and unclear pairing symmetry in Fe-based superconductors. We propose that properties of these materials should thus be measured either on fresh surfaces, e.g., on single crystals cleaved in a vacuum chamber or liquid helium or surfaces cleaned (etched) immediately before the measurement.

This work was supported by the European Commission Grant No. EC NMP3-SL-2011-283141 IRON-SEA and the Slovak Research and Development Agency under Contract No. APVV-0494-11 and DO7RP-0026-11. It is also result of the project implementation: 26240120026 supported by the Research & Development Operational Programme funded by the ERDF.